\documentclass[letterpaper,twocolumn,10pt]{article}
\usepackage{usenix-2020-09}

\definecolor{acommentColor}{rgb}{0.09, 0.45, 0.27}
\definecolor{tcommentColor}{rgb}{1.0, 0.75, 0.0}
\definecolor{ccommentColor}{rgb}{0,0.6,0.8}
\definecolor{arcommentColor}{rgb}{0.8,0.3,0}
\definecolor{rcommentColor}{rgb}{150,0,150}
\newif \ifcomments \commentstrue
\ifcomments
\newcommand{\andres}[1]{{\color{acommentColor} /* Andres: #1 */}}
\newcommand{\tom}[1]{{\color{tcommentColor} /* Tom: #1 */}}
\newcommand{\carolina}[1]{{\color{ccommentColor} /* Carolina: #1 */}}
\newcommand{\armin}[1]{{\color{arcommentColor} /* Armin: #1 */}}
\newcommand{\rachit}[1]{{\color{red} /* Rachit: #1 */}}
\newcommand{\ben}[1]{{\color{red} /* Ben: #1 */}}
\else
\newcommand{\andres}[1]{}
\newcommand{\tom}[1]{}
\newcommand{\carolina}[1]{}
\newcommand{\armin}[1]{}
\newcommand{\rachit}[1]{}
\newcommand{\ben}[1]{}
\fi

\newif\iffullversion
\fullversionfalse

\makeatletter
\renewcommand{\paragraph}{%
  \@startsection{paragraph}{4}%
  {\z@}{1.2ex \@plus 1ex \@minus .2ex}{-1em}%
  {\normalfont\normalsize\bfseries}%
}
\makeatother

\newcommand{\smidge}{{\kern .05em}}

\newcommand{\false}{\textsf{false}}

\newcommand{\true}{\mathsf{true}}

\newcommand{\bnm}{\begin{newmath}}
\newcommand{\enm}{\end{newmath}}
\newcommand{\bne}{\begin{newequation}}
\newcommand{\ene}{\end{newequation}}

\newenvironment{newmath}{\begin{displaymath}%
\setlength{\abovedisplayskip}{4pt}%
\setlength{\belowdisplayskip}{4pt}%
\setlength{\abovedisplayshortskip}{6pt}%
\setlength{\belowdisplayshortskip}{6pt} }{\end{displaymath}}

\newenvironment{newequation}{\begin{equation}%
\setlength{\abovedisplayskip}{4pt}%
\setlength{\belowdisplayskip}{4pt}%
\setlength{\abovedisplayshortskip}{6pt}%
\setlength{\belowdisplayshortskip}{6pt} }{\end{equation}}

\newcommand{\secref}[1]{Section~\ref{#1}}

\newcommand{\verylongrightarrow}[1]             
      {\setlength{\unitlength}{.01in}           
      \begin{picture}(#1,1) \put(0,0){\vector(1,0){#1}} \end{picture}}

\newlength{\saveparindent}
\newlength{\saveparskip}
\newcounter{ctr}

\newenvironment{newenum}{%
\begin{list}{{\rm (\arabic{ctr})}\hfill}{\usecounter{ctr} \labelwidth=17pt%
\labelsep=5pt \leftmargin=22pt \topsep=3pt%
\setlength{\listparindent}{\saveparindent}%
\setlength{\parsep}{\saveparskip}%
\setlength{\itemsep}{2pt} }}{\end{list}}

\definecolor{clr3_1}{RGB}{203,106,73}
\definecolor{clr3_2}{RGB}{164,108,183}
\definecolor{clr3_3}{RGB}{122,164,87}

\definecolor{clr5_1}{RGB}{75,174,141}
\definecolor{clr5_2}{RGB}{202,86,136}
\definecolor{clr5_3}{RGB}{133,160,64}
\definecolor{clr5_4}{RGB}{135,116,202}
\definecolor{clr5_5}{RGB}{202,112,64}

\definecolor{clr2_1}{RGB}{179,102,158}
\definecolor{clr2_2}{RGB}{152,152,77}

\definecolor{clr6_1}{RGB}{167,221,226}
\definecolor{clr6_2}{RGB}{230,184,179}
\definecolor{clr6_3}{RGB}{155,194,175}
\definecolor{clr6_4}{RGB}{209,187,223}
\definecolor{clr6_5}{RGB}{212,217,182}
\definecolor{clr6_6}{RGB}{170,196,226}

\definecolor{clr6_1}{RGB}{121,113,168}
\definecolor{clr6_2}{RGB}{121,177,69}
\definecolor{clr6_3}{RGB}{154,72,190}
\definecolor{clr6_4}{RGB}{89,141,108}
\definecolor{clr6_5}{RGB}{183,73,89}
\definecolor{clr6_6}{RGB}{185,124,63}

\usepackage{mathtools}
\usepackage{amsthm}
\usepackage{enumitem}
\usepackage{adjustbox}
\usepackage{setspace}
\usepackage{tikz}
\usetikzlibrary{pgfplots.groupplots, matrix,fit}
\usetikzlibrary{calc}
\usepackage{amsmath}
\usepackage{amsthm, thmtools, thm-restate}
\usepackage{amsfonts}
\usepackage{siunitx}
\usepackage{subcaption}
\usepackage{multirow}
\usepackage{xspace}
\usepackage{cryptocode}
\usepackage{amsfonts}
\usepackage{amsmath}
\usepackage{enumitem}
\usepackage{setspace}
\usepackage{adjustbox}
\usepackage{booktabs}
\usepackage{siunitx}
\usepackage{tabularx}
\usepackage{xcolor}[dvipsnames]
\usepackage{siunitx}
\usepackage{adjustbox}
\usepackage{makecell}
\usepackage{rotating}
\usepackage{xurl}

\newcommand{\enc}{\mathsf{Enc}}

\newcommand{\gzip}{\mathsf{gzip}}
\newcommand{\dup}{\mathsf{dup}}
\newcommand{\state}{\mathcal{V}}

\newcommand{\obs}{\mathcal{B}}

\newcommand{\concat}{\:\|\:}

\newcommand{\advA}{\mathcal{A}}

\begin{document}

\date{}

\title{Exploiting Leakage in Password Managers via Injection Attacks}

\author{
  {\rm Andrés Fábrega$^{1}$, Armin Namavari$^{1}$, Rachit Agarwal$^{1}$, Ben Nassi$^{2, 3}$, Thomas Ristenpart$^{1, 2}$}\\
  $^{1}$ Cornell University \hspace*{3em} $^{2}$ Cornell Tech \hspace*{3em} $^{3}$ Technion - Israel Institute of Technology
} 

\maketitle

\begin{abstract}
  This work explores \emph{injection attacks} against password
managers. In this setting, the adversary (only) controls their own application
client, which they use to ``inject'' chosen payloads to a victim's client via,
for example, sharing credentials with them. The injections are interleaved with
adversarial observations of some form of protected state (such as encrypted vault exports
or the network traffic received by the application servers), from which the
adversary backs out confidential information. We uncover a series of
general design patterns in popular password managers that lead to 
vulnerabilities allowing an adversary to efficiently recover passwords,
URLs, usernames, and attachments. We develop general attack templates to exploit
these design patterns and experimentally showcase their practical efficacy via 
analysis of ten distinct password manager applications. We disclosed our findings to these vendors, many of which deployed mitigations.

\end{abstract}

\section{Introduction}
Password-based authentication suffers from well-know pitfalls, such as the fact that users tend to choose passwords that can be easily guessed by attackers~\cite{googlepasswordsreport}. Password managers are often cited as the default solution to this problem~\cite{googlepasswordmanagers,troyhuntpasswordmanagers}, as users can offload to them the complexities of password generation, storage, and retrieval. Indeed, password managers have enjoyed a noticeable rise in popularity~\cite{googlepasswordsreport,security-org-report}, placing them among the most ubiquitous security-oriented tools.

Password managers have benefited from academic
attention~\cite{carr2020revisiting,fahl2013hey,li2014emperor,gasti2012security,silver2014password,oesch2020then,chatterjee2015cracking,petrov2022android,gilsenan2023security,golla2016security,bojinov2010kamouflage,stock2014protecting},
which has helped understand and improve their security along various dimensions.
The attacks uncovered by prior work broadly fall under two general threat
models. First are attacks that use a client-side resource controlled by
the adversary, such as a malicious
website visited by the client~\cite{li2014emperor,stock2014protecting}, a rogue application in the
victim's device~\cite{carr2020revisiting,fahl2013hey}, or the client's WiFi
network~\cite{silver2014password}. Second are adversaries that
somehow acquire a copy of a user's encrypted vault, and exploit
leakage from unencrypted vault metadata~\cite{gasti2012security,oesch2020then}
or by offline cracking attacks of a user's master
password~\cite{chatterjee2015cracking,golla2016security,bojinov2010kamouflage}.
State-of-the-art password managers are therefore designed to resist both
kinds of threats and, notably, use slow cryptographic hashing to prevent cracking
attacks for well-chosen master passwords.

In this work, we consider a new kind of threat model in which an adversary (1) controls their own application client, through which they can send chosen
payloads to the victim (for example, via the password sharing feature found in
most modern password managers); and (2) can observe some form of encrypted state
and associated metadata, such as the user's encrypted 
vault backups or network requests received by the application servers.
Borrowing terminology from prior work
in other domains (see Section~\ref{sec:related-work}), we refer to attacks in
this threat model as \emph{injection attacks}.

\begin{figure*}[t]
    \centering
    \footnotesize
    \renewcommand{\arraystretch}{1.2}
    \begin{tabular}{lllll}
     \toprule
     \textbf{Attack vector} & \textbf{Leakage} & \textbf{Adversary type} & \textbf{Vulnerable applications}\\ 
     \midrule
    Vault-health logs & Passwords & Eavesdropper & L, D, Z, K, N\\
    Vault-health logs & Passwords & Network & Z\\
    URL icon fetching & URLs & Network & D, 1P, E, R, P, N \\
    Attachment deduplication & Attachment & Eavesdropper & KX\\
    Compression & URLs and usernames & Eavesdropper & KX\\
     \bottomrule
\end{tabular}
\caption{Summary of the vulnerabilities discovered in this work, which lead to
  efficient attacks that recover sensitive information from a victim's vault. These vulnerabilities were present in ten applications we studied: LastPass (L), Dashlane (D), Zoho Vault~(Z), 1Password (1P), Enpass (E), Roboform (R), Keeper (K), NordPass (N), Proton Pass (P), and KeePassXC (KX).}
\vspace{-0.3cm}
\label{fig:attacks-summary}
\end{figure*}

The core idea behind injection attacks is that the adversary can use injections
to trigger subtle interactions in the application logic between their
data and target victim data (e.g., other passwords used by the target), 
which are reflected in their observations of ciphertexts
(e.g., inspecting their lengths) and metadata in a way that allows 
recovering sensitive information. We argue that this threat model is 
increasingly important as password managers become more complex and
feature-rich, which provides new avenues for injection mechanisms and vulnerable cross-user interactions. 

To understand whether this threat model is of practical concern or not, we
performed a security analysis of ten popular password managers that support
sharing---LastPass, Dashlane, Zoho Vault, 1Password, Enpass, Roboform, Keeper, NordPass, Proton Pass, and KeePassXC. Together
these reportedly account for over 30\% of all password manager users~\cite{security-org-report}. 
We uncover a series of exploitable vulnerabilities that implicate all
of the password managers investigated. 

Our first class of attacks exploits the fact that a common feature of password
managers is for clients to periodically log outside the device various
metrics about the ``health'' of a user's vault, such as the number of duplicate
passwords. We show how an adversary can leverage these benign-looking metrics to
perform an efficient binary-search-based dictionary attack that recovers the target user's saved
passwords. Our attacks do not require the adversary to know additional
information about the victim's saved credentials beforehand (for example, URLs
nor usernames). Five out of the ten applications are vulnerable to
this attack. In most cases, the adversary must be a passive \emph{eavesdropper}
that observes these metrics directly (for example, by
having a persistent foothold in the application servers), while for one
application the attack is feasible by a 
passive \emph{network adversary} that simply observes the HTTPS
channels under which the E2EE data is transmitted. We note that both
eavesdropping and network adversaries are within scope of the threat models
under which password managers are designed, and the ubiquity of server-side breaches~\cite{lastpass-breach,norton-breach,dropbox-breach},
combined with the difficulty of detecting such breaches~\cite{ibm-breach-report}, make it
critical that password managers resist such attacks.

Our second class of attacks exploits another feature of password
managers: clients often display a small identifying icon, such as a company
logo, alongside each of a user's saved credentials. Importantly, such icons are
only fetched once per URL, and subsequent credentials reuse the icon stored in
the client. We show how this fact allows an adversary to perform an efficient
dictionary attack on the URLs in a victim's vault. The attack always succeeds in
our experiments, and mounting it requires no additional assumptions about 
the victim's saved credentials. Six of our case study applications are vulnerable to this attack,
and in all cases exploitation only requires observations by a network adversary. 

We then turn our attention to adversaries that have an encrypted copy of the
entire vault, such as compromising a local password-protected database file or
backup of it. In this case, we analyze the security of KDBX~\cite{kdbx4}, which
is a file format used by many password managers, notably KeePass~\cite{keepass}
and its derivatives~\cite{kdbx4-projs, kdbx4}. To optimize for storage, KDBX
employs a variety of storage-saving techniques, such as file deduplication and
compression. We show two attacks exploiting these features to
recover URLs, usernames, and attachment contents. Compression and deduplication
have led to attacks against other systems before (see
Section~\ref{sec:related-work}), but our work is the first to show that these types
of vulnerabilities also arise in the context of password managers. Our attacks
target features of the underlying file format itself, and thus can potentially
be leveraged against any application that uses KDBX. We implement
a proof-of-concept for our attacks in the case of
KeePassXC, and experimentally show that its accuracy 
is sufficiently high to make it a practical threat. 

A summary of our attacks is shown in Figure~\ref{fig:attacks-summary}.
They exploit common design patterns found in password
managers, and as such other applications that employ these can be vulnerable to
our attacks. Indeed, for each of our attacks, we describe a general template for
it, which is agnostic to lower-level application details, and that can be used to
target any application that follows the relevant design pattern. More broadly, our
findings uncover higher-level issues in password manager design, and we discuss
the future work that will be required to provide generally applicable
mitigations for injection attacks.

\paragraph{Summary of contributions.} We begin the study of a new threat model
for password managers called injection attacks. We identify
  three design patterns that lead to attacks, and we implement practical attacks
  that affect a variety of applications. We stress, however, that our attack
  vectors are features of password managers, instead of bugs that are specific
  to our case study applications, and thus other password managers can
  potentially be vulnerable to them. Some of our vulnerabilities reveal new
  attack vectors, whereas others exploit known malpractices (compression and
  file deduplication). Thus, for the latter, our work is the first to show that
  these issues are also present in password managers.

Our attacks highlight broader classes of design malpractices found in password
managers (and E2EE applications more generally). We close this work by
identifying these higher-level issues, and outlining a series of takeaways for
application designers. Our results thus pave the way for future work along
various dimensions: identifying other password managers that are vulnerable to
our attacks, uncovering other patterns that lead to injection attacks, and
designing general tools to study and mitigate injection attacks. 
We expand on these ideas, and other opportunities for future work, in Section~\ref{sec:takeaways-and-mitigations}.

\paragraph{Ethics.}
All of our attacks were performed against isolated research accounts. We kept
the scale of our proof-of-concept experiments as minimal as possible, so as to
confirm the attacks' efficiency without overloading any application or cloud
servers. Some of our experiments required a high volume of network traffic, for
which we implemented local simulators that we validated via smaller experiments
with real clients.

We disclosed our findings to the four password managers
directly affected by our work, and made ourselves available
to help with mitigations before public release of our findings. Since then, many of the vendors proceeded to deploy mitigations for our reported vulnerabilities (see
Section~\ref{sec:takeaways-and-mitigations}), which are either already deployed or will be deployed soon.

\section{Related Work}\label{sec:related-work}

\paragraph{Security analysis of password managers.}
Prior work has investigated attacks on password managers in threat models
involving some combination of client-side attacks, adversarial networks, and a malicious service provider. Such examples include attacks that rely on malicious applications in the
victim device~\cite{fahl2013hey}, malicious websites that the victim is
tricked into visiting~\cite{li2014emperor}, XSS adversaries that inject
code into a website's login page~\cite{stock2014protecting}, and a rogue
WiFi network under the adversary's control~\cite{silver2014password}. The
goal of the adversary is to exploit some feature of the application, such
as password generation~\cite{oesch2020then}, autofill
policies~\cite{silver2014password}, and clipboard
vulnerabilities~\cite{carr2020revisiting, fahl2013hey}, to exfiltrate user
passwords.

More relevant to our findings, some existing attacks on password managers
involve an adversary that obtains the encrypted database of the
victim~\cite{gasti2012security,chatterjee2015cracking,golla2016security,bojinov2010kamouflage,oesch2020then}.
However, most of these attacks focus on offline cracking of the master
password, using the recovered vault as a decryption
oracle~\cite{chatterjee2015cracking,golla2016security,bojinov2010kamouflage}.
Other works, such as~\cite{gasti2012security} and~\cite{oesch2020then},
simply document unencrypted metadata in the encrypted database files
or consider a much weaker adversarial goal, namely, producing a new, valid
database after observing other valid ones (for example, by tweaking
metadata headers). Neither attack violates the confidentiality of the
password vault contents.

Our injection attack threat model is different from settings explored in
prior work on password managers. We consider an adversary that, in addition
to potentially compromising the platform and/or network, can spin up
clients of its own that interact with the victim client using standard
password manager features. The adversary uses these cross-user interactions
to mix data of its choosing with sensitive victim data. Then, via a leakage
channel, the adversary learns information about the combination of the
victim data and its own injected data.

\paragraph{Injection attacks.} Although we are the first to apply injection
attacks to password managers, prior work has studied the injection threat
model in other contexts. For
example,~\cite{cash2015leakage,xu2021searching,zhang2016all} present
attacks against searchable encryption schemes and encrypted search indexes,
where the adversary is able to inject payloads that trigger a query into
the encrypted store.
Our attacks take inspiration from techniques used in this prior work, such
as the binary search attack presented in \cite{zhang2016all}. However, our
contribution lies in how we apply these techniques through the leakage
channels and injection vectors we identify, specific to the password
manager setting.

A setting closer to ours is that of~\cite{injectionattackspaper}, which
introduces attacks against E2EE backups of messaging applications, in the
presence of an adversary that can message the victim and subsequently
observe their encrypted chat backups. A few of our attacks
(Section~\ref{sec:compromised-storage-attacks}) rely on similar attack
vectors (file deduplication and compression). However, beyond targeting
different types of applications, our attacks represent a different class of
injection attacks: our work exploits encrypted state
\emph{synchronization}, whereas~\cite{injectionattackspaper} exploits
encrypted state \emph{backups}. As such, our setting represents a richer
attack surface, with a higher frequency and granularity of observations,
and so our attacks are significantly more practical, as we explain further
in this section. Furthermore, our attacks exploit features such as health
metric logs and URL icon fetching, going beyond the attack vectors explored
in prior work on encrypted backups.

\paragraph{Attacks on compression before encryption.}
It has long been known that compression before encryption can lead to
vulnerabilities~\cite{kelsey2002compression}, which has resulted in exploits against real-world systems~\cite{rizzo2012crime, gluck2013breach, zindros2016practical, kelsey2002compression, patersonthree, hogan2023dbreach, injectionattackspaper}, such as TLS~\cite{rizzo2012crime,gluck2013breach} and the iMessage E2E encrypted messaging protocol~\cite{garman2016dancing}. Our work, however, is the first to exploit compression in the context of password managers (Section~\ref{subsec:leakage-compression}). Our attacks combine techniques from prior works (highlighted, as needed, in the attack descriptions) with novel insights in order to exploit this new setting.

Some works have studied compression in the broader context of encrypted databases, such as~\cite{patersonthree, hogan2023dbreach, injectionattackspaper}. The attacks in both~\cite{patersonthree} and~\cite{hogan2023dbreach} rely on assumptions that are not present in our setting, such as physical access to the target's handset and the ability to unlock it~\cite{patersonthree}, and little to no noise in the side channel~\cite{patersonthree, hogan2023dbreach}. Further, their attacks are tailored to the specific systems that are being targeted. The setting in~\cite{injectionattackspaper} is much more limited than our work, as the adversary is limited by daily backups, every injection results in much more noisy metadata that is added to the database, and the adversary cannot edit past injections. Thus, we devise new attacks that are significantly more practical: our attacks handle larger dictionary sizes (hundreds of items instead of, e.g., 10 to 20 items), have higher accuracy (for example, for a dictionary of size 20---the largest~\cite{injectionattackspaper} experiments with---our attacks succeed with 90\% probability instead of 20-30\%), and have an additional confirmation step to verify if the found item is the correct one or not. Further, their compression-based attacks require the victim not to send or receive external messages for multiple days, whereas our attacks run in a matter of minutes.

File deduplication, a common form of compression before encryption, has
been exploited in other contexts, such as client-side encrypted file
storage~\cite{harnik2010side,halevi2011proofs} and E2EE
messaging~\cite{injectionattackspaper}. Our injection attacks exploiting
deduplication (\secref{subsec:leakage-attach-dedup}) require new techniques
due to details of KDBX~4's architecture. In particular, KDBX~4 employs both
deduplication \emph{and} compression, which required mitigating noise from
the latter. We also note that the attack from~\cite{injectionattackspaper}
is not applicable to the applications we consider in this work.
\section{Password Managers Background}\label{sec:password-managers-architecture}
We describe the general architecture of password managers in this section.

\paragraph{Password manager abstraction.} We denote by~$U$ and~$PW$ a user of a password manager service~$S$ and their account password, respectively. The user~$U$ owns one or more devices with an application client for $S$ installed in each. The types of clients available vary by service, but these typically consist of mobile applications, desktop applications, web applications, and browsers extensions. Each client stores $U$'s saved passwords and other information in the local storage of its respective device in the form of a \emph{local vault}, which is (often) encrypted. We represent the contents of a vault by $\state = \{e_1, ..., e_m\}$, where each $e$ is a vault entry storing $U$'s credentials for a website. Each $e$ contains various fields such as username ($e_{user}$), URL ($e_{url}$), password ($e_{pw}$), and a list of attachments ($e_{attach}$).\footnote{A common feature of password managers is for users to be able to attach arbitrary files, e.g., sensitive documents, to their vault entries.}

In order to have consistent local vaults, $U$'s devices need to synchronize their state. For most password managers, this involves clients periodically communicating with platform servers, which serve as intermediaries that facilitate synchronization. To do so, clients export an E2EE version of (only) the latest state changes (e.g., new passwords added), using an application-specific symmetric encryption scheme, and send this to $S$'s servers using HTTPS. The server then forwards the updates to all other devices the next time they come online; the receiving clients decrypt the changes and update their local vaults accordingly.

Alternatively, if the service provider does not directly facilitate synchronization,~$U$ has to ensure that their clients are consistent, either by propagating updates manually, or by using out-of-band synchronization mechanisms such as storing the encrypted database file on an external cloud-storage provider. In this case, state updates are less granular, as these consist of a re-encryption and a re-upload of the entire database file to the cloud via HTTPS. Note that this setup is explicitly suggested by various applications that do not automatically synchronize clients~\cite{keepassxc-user-guide,enpass-cloud}.

State synchronization is very frequent: exports generally occur after every modification to the local database, and clients automatically check for imports (from $S$ or from modifications to their cloud-stored file) every couple of seconds; the exact periodicity varies across services.

\paragraph{Database file formats.}
Most password managers use simple data structures and file formats for their local vaults, which typically consist of encrypting each field of each $e \in \state$ separately, and sequentially organizing these entries using some lightweight file format like JSON or XML. Some file formats then additionally employ a variety of storage-saving mechanisms, such as attachment deduplication and compression, to minimize the size of the database file. This pattern is most often seen in applications that do not route information through $S$'s servers, as everything is stored on the user's device and (potentially) exported to some third-party cloud service. Conversely, other applications can leverage $S$'s servers for storage; for example, instead of storing the binary content of all attachments, local vaults can just store a pointer to a blob store managed by $S$, which contains the encrypted file itself.

\paragraph{Credential sharing.}
A common feature of modern password managers is \emph{cross-user sharing}, which allows users to jointly hold entries in their vaults. For this, whenever some user wants to share a credential~$e$ from their vault with $U$, the former derives an ephemeral key that is used to encrypt $e$, and sends this ciphertext alongside the ephemeral key (itself encrypted under a public key associated with~$U$). Next time one of~$U$'s devices comes online, its client will download the shared items and decrypt them locally to recover~$e$.

We note that, generally, $U$ must first accept the shared entry $e$ before it
gets added to their vault.\footnote{Some 
managers place additional trust constraints on password sharing: Zoho Vault only allows sharing with users who are part of the same
``organization'', which are groups that the user can belong to.} However,
subsequent updates to $e$ require no approval, and are propagated automatically
to all clients. Further, many password managers allow users to share
\emph{folders} in addition to individual entries. In this case, $U$ must once
again accept the initial share of the folder, but all future updates, including
adding or deleting new entries to the folder, require no approval.

For some applications, credential sharing is restricted to accounts
that are part of the same ``organization'', e.g., where both accounts are
affiliated with a corporate or family license for the application. In this case,
shared credentials should still provide E2EE guarantees, such as cryptographic
access control, even across all organization members. These guarantees should
hold even in the presence of privileged users such as organization
administrators (who may have access to organization-wide metadata).

\section{Threat Model and Case Studies}

While password managers have traditionally been non-interactive
applications, newer features like credential sharing require us to
re-think their security model: what attacks, if any, arise from
\emph{interaction with other malicious clients}? This question is the starting
point for our threat model.

\subsection{Threat Model}\label{subsec:threat-model}
Our threat model, which we describe in detail in this section, assumes an
adversary that (1) can inject content into the victim's vault, and (2) subsequently observe some 
form of protected application state.

\paragraph{Injections.} The injection channel of our attacks will be the aforementioned \emph{cross-user sharing} feature of password managers. That is, the adversary can share with the victim a credential $e$, which gets incorporated and synchronized across all victim vaults, resulting in $e$ being ``injected'' into~$\state$. We stress that, in our setting, the adversary exclusively controls an (unmodified) application client, and sends their payloads through the standard interface provided by the application. Thus, tampering with either the victim or the adversary's clients, or controlling any other parts of the environment, is explicitly out of scope.

As mentioned earlier, password managers often require the victim to first accept the shared items before they are incorporated into their vault, which is thus a necessary assumption required to establish an injection channel. We stress, however, that all of our attacks require the victim to accept a \emph{single} shared item, as all subsequent updates to it, through which injections are performed, require no approval. As such, our threat model assumes some initial degree of trust between the user and the adversary, or that the latter can trick the former into accepting a share request. Accepting such a request, however, should not lead to disclosure of a user's vault content.

\paragraph{Observations.} Our threat model then assumes that the adversary has \emph{persistent access} to some function of the data that leaves the victim's device, e.g., as a result of recurring backups or state synchronization between devices. We distinguish between two variants, depending on the trust assumptions required for each attack: (1) an \emph{eavesdropping adversary}, who has access to the E2EE data itself, e.g., ciphertexts of new passwords, as well as other plaintext metadata; and (2) a weaker \emph{network adversary} that can only observe the HTTPS packets under which the E2EE data is transmitted. We highlight the difference between both settings in Figure~\ref{fig:adv-variants}. We discuss the periodicity of synchronization (and, hence, the frequency of adversarial observations) for all case study applications in the relevant attack sections.

\definecolor{color1}{HTML}{009E73}
\definecolor{color2}{HTML}{E69F00}
\begin{figure}[t]
    \centering
    \includegraphics[width=\columnwidth]{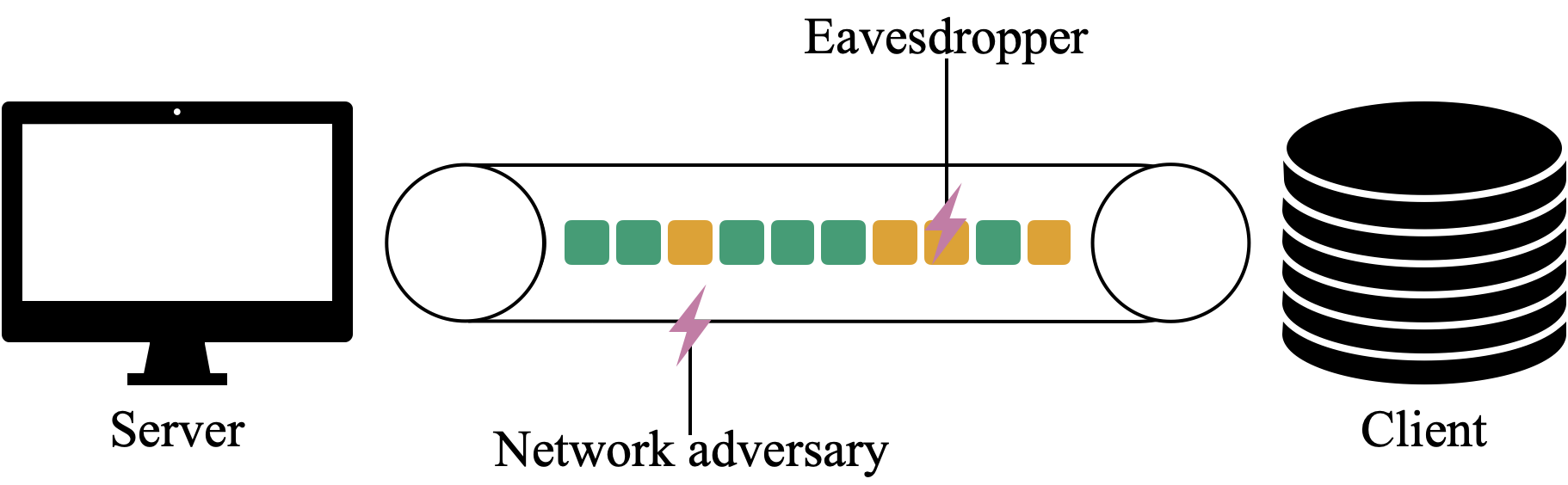}
    \caption{A network adversary can observe the HTTPS packets under which a mix of E2EE data (\textcolor{color1}{green} squares) and plaintext metadata (\textcolor{color2}{orange} squares) is transmitted; an eavesdropper has direct access to these.}
    \label{fig:adv-variants}
\end{figure}

Password managers promise end-to-end confidentiality~\cite{lastpasse2ee, bitwardene2ee,1passworde2ee}, i.e., a user's data is compromised only if their master password is leaked, and so password managers are designed to protect against eavesdropping and network adversaries. A network adversary can arise from any number of traffic analysis techniques, such as ARP or DNS spoofing, BGP hijacking, router compromise, a malicious ISP, etc. An eavesdropping adversary generally corresponds to a malicious or breached service or cloud provider, or from a privileged user within an organization with access to metadata (e.g., an administrator). However, they may also arise from other attacks against the TLS layer of traffic, e.g., attacks on certificate authorities (CA) or malicious client-side proxies. Breaches on password managers and cloud services~\cite{lastpass-breach,norton-breach,dropbox-breach} suggest that such an eavesdropping adversary is a realistic threat model, and indeed \emph{persistent} access is a practical concern: an IBM report from 2022~\cite{ibm-breach-report} found that the average time it takes to identify a breach is 212 days.

The process of injections and (passive) observations occurs iteratively: the adversary sends a payload to the victim (either by sharing a new credential, or modifying older ones), observes the resulting encrypted state—--which is now a function of both adversarial and sensitive data---adaptatively chooses the next payload to inject, and so on. The adversary's goal is then to back out confidential information from their observations.

A priori, no sensitive user data should be leaked, since the victim's confidential information is encrypted before leaving their device. However, our attacks exploit the fact that application logic often \emph{mixes} personal and externally-received data in subtle ways which may leak information. Thus, an adversary can use injections to trigger these \emph{cross-user interactions}, which are then reflected in the exported application state, revealing confidential data. We stress that injections attacks assume strong cryptographic primitives, and thus low-level, cryptanalytic vulnerabilities are outside their scope; instead, the \emph{lengths} of ciphertexts and plaintext metadata form the basis of our~attacks.

In addition to the adversary's injections, the state changes between
observations may additionally contain new, benign data added by the victim while
the attack is running; we refer to this as the ``noisy device'' setting, and to
the case where the only additions are the adversary's injections as the ``quiet
device'' setting. Some attacks are robust to noise, while others require the
victim's client to be quiet while the attack is running; we will specify the
noise assumptions required for each attack in the relevant sections.

\paragraph{Out-of-scope attacks.}
  
We do not consider  attacks in which
  a compromised service attempts to deploy client code 
    containing backdoors. All E2EE threat models implicitly or explicitly assume trusted client
    software. Improved assurance here can be aided via mechanisms such as 
    public auditability of software or monitoring via binary transparency
    services (e.g.,~\cite{newman2022sigstore}). Without trustworthy client side
    software, no confidentiality is possible. More pragmatically, we note that
    such client subversion would require an active adversary
    that hijacks client code distribution, while our injection attacks would be
    easier to mount, requiring only
    a passive adversary that either compromises some platform server (e.g., a
    web or storage server) to enable
    eavesdropping or has visibility into the target's network communications.

    In our threat model, we assume that the adversarially controlled client that
    performs injections honestly follows the protocol. One could also  
    consider a fully malicious client that deviates from
    the E2EE protocol when performing injections. 
    We are unaware of any additional attacks that this would enable.

\subsection{A Corpus of Password Managers}

  We gathered a set of password managers to experimentally investigate the
  feasibility of injection attacks. To build a list of targets, our starting
  point was a report by Security.org~\cite{security-org-report} that includes a
  list of password managers ranked by popularity. We also considered informal
  online polls on password manager popularity~\cite{reddit-poll,reddit-comparison-apps}.  
  We investigated the advertised features of each of the resulting 16 applications, and
  excluded those that do not meet either of two inclusion criteria: (1) the
  application must support cross-user credential sharing,
  and (2)~the application must target cryptographic access control for
  shared credentials. The first requirement rules out applications that are
  unlikely to have any way to inject adversarial content into a target, and the
  second requirement rules out applications for which simpler attacks would
  work in our threat model, due to lack of E2EE security guarantees. 

  All browser-integrated password managers were excluded due to the first
  criteria, as they 
  do not yet support credential sharing. The second criteria ruled out one
  application, Bitwarden, which does support credential sharing, but not in a
  cryptographically secure way:
  Bitwarden has
  support for establishing separate ``collections'' within an organization, but
  surprisingly all collections use the same secret key. So, in our threat model,
  any organization member can already decrypt all organization
  credentials.\footnote{We
  disclosed this to Bitwarden, who confirmed that privilege separation
  across collections is ``at the authorization level, not at the encryption
  level''. Thus, they target a weaker security model than other applications.}

    Our final list
    of applications consisted of LastPass (v4.123), Dashlane (v6.2346), Zoho Vault (v3.8), 1Password (v2.25.0),
    Enpass (v6.11.0), Roboform (v9.6.2), Keeper (v116.18.0), NordPass (v.5.16), Proton Pass (v1.17.4), and KeePassXC (v2.7.6). These
    applications cover over 30\% of all password manager
    users according to~\cite{security-org-report}, and all follow the basic abstraction
    presented in Section~\ref{sec:password-managers-architecture}. All application except for KeePassXC and Enpass rely on the
    application servers for stateful storage and synchronization. These tend to
    use simple file formats for their local vaults (see~\cite{oesch2020then}
    for an overview), with no additional storage-saving mechanisms. Conversely,
    KeePassXC and Enpass do not natively handle synchronizing across clients. Instead, they
    suggest~\cite{keepassxc-user-guide, enpass-cloud} storing the database file
    in an external cloud provider to which all user devices have access.
    
    KeePassXC's file format, KDBX~\cite{kdbx4} (detailed in
    Section~\ref{sec:compromised-storage-attacks}), supports attachment
    deduplication and vault compression. 
    We note that KDBX is used by other password managers, which are ports and
    derivatives of KeePass~\cite{keepass}, a popular open-source password
    manager that introduced this file format. See~\cite{kdbx4-projs, kdbx4} for
    comprehensive lists of applications that use KDBX. Even though we implement
    our attacks on one such example, KeePassXC (which is a more cross-platform
    and feature rich version of the original KeePass), our attacks target the
    underlying file format itself, and thus other applications that use it may
    be vulnerable to our attacks.

\subsection{Overview of Attacks}
\definecolor{color1}{HTML}{117733}
\definecolor{color2}{HTML}{882255}
\begin{figure*}[t]
    \centering
    \scriptsize
    \renewcommand{\arraystretch}{1}
    \begin{tabular}{ll|ccccccccccc}
      \textbf{Attack} & \textbf{Pre-conditions} & \begin{rotate}{50} \textbf{LastPass} \end{rotate} & \begin{rotate}{50} \textbf{Dashlane} \end{rotate}& \begin{rotate}{50} \textbf{Zoho Vault} \end{rotate}& \begin{rotate}{50} \textbf{1Password} \end{rotate}& \begin{rotate}{50} \textbf{Enpass} \end{rotate}& \begin{rotate}{50} \textbf{Roboform} \end{rotate}& \begin{rotate}{50} \textbf{Keeper} \end{rotate}& \begin{rotate}{50} \textbf{NordPass} \end{rotate}& \begin{rotate}{50} \textbf{Proton Pass} \end{rotate}& \begin{rotate}{50} \textbf{KeePassXC} \end{rotate}\\ 
     \midrule
     \multirow{3}{*}{App-wide metrics (Section~\ref{sec:leakage-from-security-metrics})}  & Support for reports of duplicate password & \textcolor{color1}{\checkmark} & \textcolor{color1}{\checkmark} & \textcolor{color1}{\checkmark} & \textcolor{color1}{\checkmark} & \textcolor{color1}{\checkmark} & \textcolor{color1}{\checkmark} & \textcolor{color1}{\checkmark} & \textcolor{color1}{\checkmark} & \textcolor{color1}{\checkmark} & \textcolor{color1}{\checkmark} \\
     & No. of duplicates computed across all passwords & \textcolor{color1}{\checkmark} & \textcolor{color1}{\checkmark} & \textcolor{color1}{\checkmark} & \textcolor{color2}{$\times$} & \textcolor{color2}{$\times$} & \textcolor{color2}{$\times$} & \textcolor{color1}{\checkmark} & \textcolor{color1}{\checkmark} & \textcolor{color1}{\checkmark} & \textcolor{color1}{\checkmark} \\
     &  No. of duplicates logged outside the device & \textcolor{color1}{\checkmark} & \textcolor{color1}{\checkmark} & \textcolor{color1}{\checkmark} & \textcolor{color1}{\checkmark} & \textcolor{color1}{\checkmark} & \textcolor{color1}{\checkmark} & \textcolor{color1}{\checkmark} & \textcolor{color1}{\checkmark} & \textcolor{color2}{$\times$} & \textcolor{color2}{$\times$}\\
    \midrule
     \multirow{3}{*}{URL icon fetching (Section~\ref{sec:leake-url-icons})} & Support for URL icons & \textcolor{color1}{\checkmark} & \textcolor{color1}{\checkmark} & \textcolor{color1}{\checkmark} & \textcolor{color1}{\checkmark} & \textcolor{color1}{\checkmark} & \textcolor{color1}{\checkmark} & \textcolor{color1}{\checkmark} & \textcolor{color1}{\checkmark} & \textcolor{color1}{\checkmark} & \textcolor{color1}{\checkmark} \\
    & URL icons fetched from servers & \textcolor{color1}{\checkmark} & \textcolor{color1}{\checkmark} & \textcolor{color2}{$\times$} & \textcolor{color1}{\checkmark} & \textcolor{color1}{\checkmark} & \textcolor{color1}{\checkmark} & \textcolor{color2}{$\times$} & \textcolor{color1}{\checkmark} & \textcolor{color1}{\checkmark} & \textcolor{color1}{\checkmark} \\
    & Fetched URL icons re-used across all vault items & \textcolor{color2}{$\times$} &  \textcolor{color1}{\checkmark} & \textcolor{color2}{$\times$} &  \textcolor{color1}{\checkmark} &  \textcolor{color1}{\checkmark} &  \textcolor{color1}{\checkmark} & \textcolor{color2}{$\times$} & \textcolor{color1}{\checkmark} & \textcolor{color1}{\checkmark} & \textcolor{color2}{$\times$}\\
    \midrule
\end{tabular}
\caption{Summary of the pre-conditions required for an application to be vulnerable for two of our attacks, each represented by a set of rows, and the conditions satisfied by each application we analyzed. An application that has a \textcolor{color1}{\checkmark} for all pre-conditions in a set of rows is thus vulnerable to the attack.}
\vspace{-0.3cm}
\label{fig:attacks-preconditions}
\end{figure*}

Our security analysis of the case study applications uncovered three main classes of attacks in the context of injection attacks. We provide an overview of these here, and discuss them in detail in the next three sections.

    \paragraph{Attack \#1: application-wide metrics (Section~\ref{sec:leakage-from-security-metrics}).} The first attack arises from the fact that many password managers compute sensitive metrics, such as the number of duplicate passwords in the vault, across both personal \emph{and} shared vault entries. If these metrics are logged somewhere outside the device (e.g., the application servers), an adversary can trigger fluctuations in these metrics with injections, and observe how they are updated in the external location. As such, besides injecting credentials, in this attack the adversary needs to have access to the location where the metrics are logged (e.g, a foothold in the application servers.)

    \paragraph{Attack \#2: URL icon fetching (Section~\ref{sec:leake-url-icons}).} The second attack arises from the fact that many password managers display a graphical icon next to each credential, identifying the website for it. This icon is only fetched once from the application servers, and future entries for the same website reuse the image that is stored in the client. Thus, an adversary can use this to test whether the victim has a credential for a particular URL or not, depending on whether the victim's client re-fetches the icon of an injected URL. So, besides updating credentials, in this attack the adversary needs to be able to observe the HTTPS requests that leave the victim's client, or have a foothold in the server where the icons are fetched from.

    \paragraph{Attack \#3: storage-saving mechanisms (Section~\ref{sec:compromised-storage-attacks}).} The third attacks arise from storage-saving mechanisms that an application may use to decrease the size of the encrypted vault. At a high-level, since these mechanisms remove redundancy in the plaintext vault, the adversary can tell whether their content matched the victim's other credentials or not by looking at fluctuations in the size of the encrypted vault. So, besides injecting credentials, in this attack the adversary would need to have persistent access to the victim's encrypted vault (for example, in a cloud provider where the victim uploads backups of their vault). In our work, we show attacks against two such mechanisms: database compression and attachment deduplication.

    In Table~\ref{fig:attacks-preconditions}, we summarize the main pre-conditions required for an application to be vulnerable to our first two attacks, as well as the pre-conditions satisfied by each application in our study. We leave implicit the fact that an application must satisfy the inclusion criteria for our threat model to begin with, e.g., support credential sharing. For our compression-based attack, the only pre-condition is that the application compresses the database file across all vault entries; for our deduplication based attack, the preconditions are (1) the application has support for attachment deduplication, and (2) deduplication occurs across attachments in all vault entries, irrespective of sender. From the applications we investigated, only KeePassXC meets the pre-conditions for these two attacks.

\section{Attacks From Application-wide Metrics}\label{sec:leakage-from-security-metrics}
Equipped with the background from the preceding sections, we proceed to describe our attacks, starting with our first class of attacks in this section.

Our close study of the network traffic of various password managers revealed that many of them periodically log outside the user's device a variety of metrics about the overall ``health'' of a user's vault, most notably the number of reused passwords. These logs are sent to either the application servers directly, or to some other member of the user's organization, e.g., an administrator with access to a company-wide security dashboard. These metrics, however, are computed across both personal \emph{and} shared vault entries. As we will show, the adversary can leverage this fact to perform an efficient dictionary attack on a user's passwords.

Dictionary attacks are a standard adversarial goal in the context of password-based authentication, as it is common for users to choose weak or reused passwords~\cite{googlepasswordsreport}. Importantly, and perhaps surprisingly, this is a problem even among users of password managers~\cite{pearman2019people,lyastani2018better}. This arises from, for example, users that import but do not update old passwords; or users that use password managers for convenience features (e.g., autofill) rather than for security features (e.g., password generation), which is a common practice~\cite{fagan2017investigation}.

We now proceed to describe the attack. We first explain the general structure of it, which serves as an attack template that can be used against any password manager that satisfies the pre-conditions for the attack. For now, we will assume that the adversary is an eavesdropper, and has a persistent foothold to wherever the vault metrics are stored, as described in Section~\ref{subsec:threat-model}

\paragraph{Attack description.} Given a dictionary of candidate passwords $P \coloneqq  \{p_1, ..., p_n\}$, the adversary $\advA$ wants to determine which $p_i \in P$ is present in $U$'s vault $
    \state$. Let $\dup(\state)$ be the number of duplicate passwords across all entries in $U$'s vault, i.e.,
$\dup(\state) = \frac{1}{2}\sum_{e', e'' \in \state} (e'_{pw} = e''_{pw})$.
In this setting, we assume~$U$'s clients periodically send $\dup(\state)$ to some external location (e.g., automatically every couple of seconds or as a result of a particular action). The key idea behind this vulnerability is that, since $\dup(\state)$ is computed across both personal and shared passwords, an eavesdropping adversary with access to this metric can use it as an oracle to determine whether a candidate password $p$ is in the victim's vault or not. Namely, they can compare the value of $\dup(\state)$ before and after injecting an entry $e$ such that $e_{pw} = p$; if $\dup(\state)$ stays the same, it must be the case that $p$ is not yet in $\state$, and thus is not one of the personal passwords of $U$. Conversely, $\dup(\state)$ increases if and only if $p$ matches one of $U$'s personal passwords.

The adversary can leverage this oracle to find all passwords in $P$ that are in $U$'s vault. There are two injection strategies that $\advA$ can use. If the application supports shared folders, instead of ``querying'' the oracle with each candidate password, one at a time, $\advA$ can use a binary-search injection strategy to arrive at the target password more quickly, as follows:

\begin{newenum}
    \item \emph{Setup:} establish a shared folder $F$ with $U$, which requires $U$ to accept the share.
    \item \emph{Baseline measurement:} wait for a network request from $U$ that contains the initial $n' = \dup(\state)$.
    \item \emph{Inject:} split $P$ into two halves $P', P''$ of equal size. For each $p \in P'$, create an entry $e$ in $F$ such that $e_{pw} = p$.
    \item \emph{Measure:} wait for a network request from $U$ with an updated $n'' = \dup(\state)$.
    \item \emph{Recurse:} set $P_{found} = P'$ if $n'' > n'$, and $P_{found} = P''$ if $n'' = n'$. If $|P_{found}| = 1$, output it as the target password. Otherwise, repeat steps 3--4 with $P = P_{found}$ as the input set and $n''$ as the baseline measurement.
\end{newenum}

If the application instead only supports sharing individual entries, $\advA$ must use a slower, sequential injection strategy. Let $t$ be the number of individually-shared entries between $\advA$ and $U$, where $t < \frac{n}{2}$ (otherwise, $\advA$ can just use the binary-search strategy from above). In this case, $\advA$ can inject $t$ candidate passwords at a time, interleaved between observations of $\dup(\state)$, by updating the password field in each of the shared entries with a new batch of $t$ candidate items. Eventually, one of these injections will trigger an increase in $\dup(\state)$. Then, to find which of the $t$ items of the prior injection is actually the reused password, $\advA$ can run the binary-search attack from before, using the $t$ shared entries as a ``folder''.

Since $\dup(\state)$ is deterministic, both variants of the attack always find the target password, if present, or confirm that no password is in the vault. The binary-search injection strategy requires $\lceil \log_2 n \rceil$ injections and observing the same number of network requests containing $\dup(\state)$; the sequential injection strategy instead requires $\frac{n}{t} + \lceil \log_2 t \rceil$ injections and observations. Which strategy to use depends on the target application, and whether it supports shared folders or not. Further, as we discuss in Section~\ref{subsec:case-studies-leakage-from-security-metrics}, in some cases $\dup(\state)$ reveals additional information, which allows $\advA$ to speed up our generic injection strategies even further.

The exact wall-clock time of the attack depends, of course, on how often these requests are triggered, which varies across applications (we discuss some examples in Section~\ref{subsec:case-studies-leakage-from-security-metrics}). Note that the runtime of the attack is independent of the number of user accounts in $\state$. That is, since $\dup(\state)$ is computed across all accounts, our attack essentially checks each password on \emph{all} user websites at once. Further, our attack can be easily tweaked to find all matching passwords, instead of just one, by recursing into all branches that increase $\dup(\state)$. In addition, our attack confirms to $\advA$ if no password in $P$ is in $\state$, which would be reflected by the fact that $\dup(\state)$ does not increase after all candidate items have been injected.

\subsection{Vulnerable Password Managers}\label{subsec:case-studies-leakage-from-security-metrics}
In summary, the pre-conditions that an application must satisfy to be vulnerable to our generic attack are: (1) have support for vault-health metrics that contain the number of duplicate passwords; (2) the number of duplicate passwords must be computed across all vault entries; and (3) these metrics must be logged somewhere outside the victim's device (for example, in the application servers or in some organization administrator portal). We performed an analysis of all ten applications and found that LastPass, Dashlane, Zoho Vault, Keeper, and NordPass all satisfy these pre-conditions, and are thus vulnerable to our attack. We show in Figure~\ref{fig:attacks-preconditions} the breakdown of pre-conditions across all ten~applications.
    
To experimentally validate our attack, we deployed proof-of-concept implementations of it against three of these vulnerable applications---LastPass, Dashlane, and Zoho Vault. We discuss these below, and refer readers to Appendix~\ref{sec:leakage-metrics-attacks-details} for additional details of these.

In all three applications, updates to shared credentials are automatically synchronized across all clients with access to these, i.e., adversarial injections are automatically incorporated into $\state$. The main difference between each application is then the frequency of when clients log $\dup(\state)$ to the applications servers. In LastPass, logs of $\dup(\state)$ are triggered automatically after a client imports the updates to the shared credentials. In Dashlane and Zoho Vault, they are triggered once per day. A second notable difference is that LastPass supports shared folders, but Dashlane does not. Thus, $\advA$ can use the binary-search and sequential injection strategies for the former and the latter, respectively. For Zoho Vault, the generic binary-search injection strategy is also feasible. However, in fact, application-specific details allow $\advA$ to significantly speed up the attack, allowing them to recover \emph{all} matching passwords with a \emph{single} injection. Zoho Vault's vault-health logs consist of an array $O$ of JSON objects, such that there is one such object for each entry $e \in \state$, containing metadata about the corresponding entry. In particular, each object contains a ``$\texttt{reused}$'' field, which maps to a boolean indicating whether $e_{pw}$ is a duplicate or not. So, $\advA$ can inject all candidate passwords in a single batch, scan $O$ next time the logs are triggered, and identify whether the objects in $O$ corresponding to the injected credentials are marked as a duplicate or not, thus revealing if that candidate password was already present in the user's vault. We discuss this optimized attack in more detail in more detail in Appendix~\ref{sec:leakage-metrics-attacks-details}.

In all three applications, no interaction is required from~$U$, and the attack only requires~$U$ to be logged into their account (for example, in their Chrome extension or web vault). Importantly, the victim need not have their vault open while the attack runs: if a user is logged in, background scripts check for updates, sync them to the vault, and trigger the requests to $\dup(\state)$. As such, the attack can operate silently in the background while $U$ is engaged in other tasks. Further, the attacks do not require the victim client to be quiet while the attack is running, and thus fall under the noisy device setting. So, $\advA$ can just silently run an extended attack if need be, irrespective of the activity of the victim, until the passwords are found.

\paragraph{Attack implementation.} We used an open-source list of common passwords~\cite{githubcommonpasswords} to gather our testing data, and confirmed that our attacks successfully recover the target string every time. Since our attacks are false-positive free---in particular, their correctness does not depend on the distribution nor number of candidate passwords---we only tested our attacks on small, proof-of-concept workloads, and thus confirmed the correctness of the attack without overloading the applications' servers.

For each application, our experimental setup consisted of two test accounts on two different devices, representing the victim and the adversary, with a shared folder between them. The list of candidate passwords consisted of a subset of the aforementioned list of common passwords. We sampled uniformly at random a password from the list of candidate passwords, added it to the victim's vault from their machine, and ran the attacks from the adversary's machine. We set up a proxy server in the victim's machine to capture all outgoing network requests, and inspect the traffic from the victim to the server during the attacks. This simulates the information that would be learned by a malicious server. We explain additional details of our testing methodology in Appendix~\ref{sec:methodology}.

\paragraph{Extension to a network adversary.} For some applications, our attack may be able to be modified to assume a weaker network adversary instead of an eavesdropper. The key challenge is that a network adversary does not have access to the plaintext value of $\dup(\state)$, as this is transmitted through encrypted channels. As such, we require an additional property from applications: that changes in the number of duplicate passwords result in changes in the size of the payload of the HTTPS requests that transmits the vault-health logs.\footnote{We also technically require that request bodies are not compressed, but this is almost always true~\cite{http-request-no-compression} (including for Zoho Vault).}

From the applications vulnerable to an eavesdropping adversary, only Zoho Vault can be extended to a network adversary. Since there is a separate boolean string in $O$ for each vault entry, $\advA$ can leverage the subtle fact that the string $\false$ has one additional character than the string $\true$, and thus the payload of the HTTPS request with $O$ as its payload will fluctuate by $1$ byte depending on whether each password is a duplicate or not. As such, our binary-search injection strategy can be used once again, by determining whether a batch of injections contains the target password or not based on this 1-byte difference. We discuss this refinement in more detail in Appendix~\ref{sec:leakage-metrics-attacks-details}.
\section{Attacks from URL Icon Fetching}\label{sec:leake-url-icons}
A feature of some password managers is to display a small identifying icon, such as a company logo, next to each vault entry. To do so, clients send a request to the server with the URL for the website, in plaintext, and receive back an image file with the icon. This clearly leaks the victim's URLs to an eavesdropping adversary (prior work~\cite{oesch2020then} has pointed this out already for a few applications). However, as we show in this section, these requests can also be leveraged by a weaker \emph{network} adversary to leak information about a victim's URLs.

This leakage arises from the fact that URL icons are fetched only \emph{once}: any new entries for websites for which there is already a credential simply reuse the icon that is already on the client, and no new fetch request is sent to the server. These requests thus serve as oracle to determine whether a candidate URL $w$ is in the victim's vault or not: $\advA$ can inject an entry $e$ such that $e_{url} = w$, wait for $U$'s client to (automatically) synchronize this new credential, and observe if an icon fetch request is triggered or not; the latter indicates that $w$ is already present in $U$'s vault, since $e$ is reusing the URL icon that is already downloaded. Importantly, note that this attack only requires the adversary to know whether a request to fetch the icon is \emph{triggered or not}, and that, unlike our first attack, the actual (plaintext) payload of the request is not relevant. This information is also visible to a network adversary, who simply monitors whether a fetch request is triggered or not. Note, however, that the adversary cannot directly tell whether the HTTPS requests to the endpoint where icons are fetched correspond to their injected credentials or not, and thus our attack requires the assumption that the victim is not adding or modifying credentials while the attack is running, as this could lead to false negatives.

Analogous to our attack from Section~\ref{sec:leakage-from-security-metrics}, this basic oracle can be used to perform a dictionary attack on the websites stored on the victim's vault. In particular, the same sequential and binary-search injection strategies can be used, depending on whether the application supports shared folders or not. 

\subsection{Vulnerable Password Managers}
In summary, the pre-conditions that an application must satisfy to be vulnerable to our generic attack are: (1) have support for URL icons; (2) fetch these URL icons from the application servers (instead of, e.g., storing them all in the client to begin with); and (3) re-use the stored icons across all vault entries. Our analysis of all ten applications revealed that Dashlane, 1Password, Enpass, Roboform, NordPass, and Proton Pass all satisfy these pre-conditions. We show in Figure~\ref{fig:attacks-preconditions} the breakdown of pre-conditions across all ten applications.

We experimentally validated our attack against Dashlane, using the Majestic ranking of top websites~\cite{majestic-list} to gather our testing data, using the same experimental setup as in Section~\ref{sec:leakage-from-security-metrics}, and confirmed that our attacks successfully recovers the target URL. We discuss additional details of Dashlane in Appendix~\ref{sec:leakage-url-icons-details}. Note that, as before, the correctness of the attack does not depend on the distribution nor number of candidate URLs, since icon fetch requests are deterministic, and as such it was sufficient to test on small workloads that did not overload their application servers.
\section{Attacks from Storage-Saving Mechanisms}\label{sec:compromised-storage-attacks}
As discussed in Section~\ref{sec:password-managers-architecture}, some password managers employ a variety of storage-saving techniques to reduce the size of their encrypted vault files. In this section, we show how two such mechanisms---file compression and attachment deduplication---lead to injection attacks against password managers. These have both been studied in other domains before (see Section~\ref{sec:related-work}), and here we show that they also lead to exploits in the context of password managers. 

From the ten applications of our study, only KeePassXC supported storage-saving mechanisms. This is due to the fact that deduplication and compression are part of the specification of its underlying file format, KDBX~4, which is used by a variety of other password managers. Even though we implement our attacks on KeePassXC, we stress that our vulnerabilities target KDBX~4 itself, and thus any application that uses it (and that supports credential sharing) can be vulnerable to them.

Looking ahead, our attacks rely on the adversary being able to observe fluctuations in the \emph{size} of the encrypted database file. Note that, if an application relies on third-party cloud providers to synchronize devices, a network adversary is sufficient: since each update re-uploads the entire encrypted file, the size of the payload of the HTTPS request can be used to detect fluctuations in the underlying file size. Further, since the attacks rely on very precise measurements, they operate in the quiet device setting, i.e., we assume the victim does not interact with their vault while the attack is running. However, our attacks benefit from the fact that modifications to shared entries are propagated automatically to the victim's vault, and thus our attacks can be run in a few minutes.

\paragraph{Background on KDBX~4.} Broadly speaking, a database~$\state$ serialized in KDBX~4 file format (the latest version of KDBX) consists of three main parts: an outer header~$H_{out}$, an inner header~$H_{in}$, and an XML payload~$I$.

The file starts with~$H_{out}$, which is unencrypted, and contains metadata about the database, such as cipher information, KDF parameters, whether the XML payload is compressed or not, etc. This is followed immediately by~$H_{in}$ and~$I$ (explained below), which are both under a single layer of encryption---using one of AES-CBC, Twofish-CBC, or ChaCha20---with a bespoke authentication mechanism; the details of the latter are unimportant for our attacks, so we omit them hereafter. Users can select their preferred encryption cipher in the database settings, with the default being ChaCha20. As usual, the encryption key~$K$ used is derived from $PW$, using either AES-KDF or Argon2. Then,~$H_{in}$ and~$I$ may optionally be compressed with gzip~\cite{deutsch1996gzip} before being encrypted, which is also a tunable parameter that is on by default.

The inner header $H_{in}$ consists of two main parts: the encryption key~$K'$ used in the second encryption layer (more on this later), and an array $A$ storing the binary content of all attachments across all entries in~$\state$ concatenated together. Importantly, only one copy of each binary file is stored in~$A$, even if the attachment is added multiple times to the database.

The XML payload~$I$ contains the user's data itself (excluding attachments, which are in~$H_{in}$). For every entry~$e \in \state$, there is a corresponding XML element in the payload, where each field of~$e$ is saved as an XLM subelement. There are two important types of subelements: (1) for every attachment in the list~$e_{attach}$, if any, there will be a subelement with an integer denoting the index within~$A$ corresponding to the binary content of this attachment; (2) the subelement for~$e_{pw}$, which stores an encrypted copy of~$e_{pw}$ using ChaCha20 and~$K'$. There is no authentication on this inner encryption layer, since it is already under the authentication mechanism of the outer one. Further, the key $K'$ is rotated after every modification to the database, and all passwords are re-encrypted. Other fields of~$e$, such as~$e_{user}$ and~$e_{url}$, are stored as unencrypted subelements; we denote by~$\hat{e}$ all subelements of~$e$, excluding~$e_{pw}$, concatenated together. 

Putting it all together, the serialization of $\state \coloneqq (e_1, ..., e_m)$ has the following structure:
\[\obs \coloneqq H_{out} \concat \enc_K\Bigl(\gzip\bigl(H_{in} \concat I\bigr)\Bigr)\]
where $H_{in}$ and $I$ have the following form:
\[H_{in} \concat I = K' \concat A \concat \enc_{K'}(e_{1, pw}) \concat \hat{e_1} \concat ... \concat \enc_{K'}(e_{m, pw}) \concat \hat{e_m}\]

For simplicity, this notation omits low-level details such as XML tags and metadata that are not relevant to our attacks. We refer readers to documentation such as~\cite{kdbx4} for a more complete treatment.

Our attacks target the underlying file format itself, and any password manager that uses it may be vulnerable to them. Our attacks, however, require two assumptions regarding the target application and its users: that the application has support for cross-user sharing, and that compression is left on (note that official documentation from KeePass, the designers of KDBX, explicitly state that ``it is not recommended to save databases without compression''~\cite{kdbx4-db-settings}.) Further, for clarity of presentation, we will assume that ChaCha20 is indeed the cipher used to encrypt the outer encryption layer, since it is the only stream cipher out of the three supported ones, but note that our attacks can be modified to work with the other~ciphers.

The application on which we implement our attacks is KeePassXC, which is a newer, feature-rich port of the original KeePass. KeePassXC supports cross-user sharing (both for individual entries and folders) which are added as new XML elements in~$I$, just like personal entries. Updates to these shared credentials are synchronized automatically if the victim has their vault unlocked (e.g., open in the background), and there is just a short time delay of 1-2 seconds.

\subsection{Leakage from Attachment Deduplication}\label{subsec:leakage-attach-dedup}
As discussed earlier, the KDBX~4 file format stores only one copy of the binary content of each attachment in $A$, even if it is received multiple times: whenever $U$ adds a new attachment to their vault, it is compared against all attachments in $A$ (by computing its checksum using a hash function), and added if and only if there is no match. If there is a match, the attachment pointer in the XML element for this entry simply refers back to the index of the first copy in $A$. Notably, deduplication occurs across both personal \emph{and} shared attachments, which leads to a cross-user interaction that an adversary can leverage to leak information about the user's saved attachments.

\paragraph{Attack description.} Our attack consists of a dictionary attack on attachments. That is, for a list of candidate attachments $W \coloneqq \{w_1, ..., w_n\}$, the adversary wants to determine which $w_i$, if any, is in $\state$. We first note that there is a straightforward (but inefficient) attack. Assume that $U$ and $\advA$ have a shared entry $e$. Then, $\advA$ can simply add each $w \in W$ to $e_{attach}$, one at a time, until some attachment does not increase the size of $\obs$, which implies that it got deduplicated and thus is present in $\state$. This naive injection strategy is false-positive free, since deduplication is deterministic.\footnote{The only edge case in which this attack may fail is if the candidate files are very small, such as less than 50 bytes. This is due to the fact that minimal noise is generated by the re-encryption of passwords after every save. In this case, however, $\advA$ can simply inject each $w_i$ multiple times, and compute the average file size, which mitigates the noise.} However, the number of injections scales linearly with $|W|$. We now describe a more complex injection strategy that an adversary can use for larger $|W|$. This second strategy leads to more false positives, and as such there is a trade-off between success rate and number of required injections.

Our refined strategy consists of a binary-search attack, which requires only $\lceil \log_2 n \rceil$ injections, analogous to that of Section~\ref{sec:leakage-from-security-metrics}: instead of injecting each attachment, one at a time, $\advA$ can split $W$ into two sets $W'$ and $W''$ of equal size, and inject \emph{all} attachments in $W'$ in a single batch, followed by all attachments in $W''$ in a second one. Then, one of these injections will increase $|\obs|$ by less than $|W^i|$, which means that some attachment in it got deduplicated. We can then recurse into this set, and repeat this attack iteratively until the target attachment is found.

A challenge with this approach is that, recall, the list of attachments is gzip-compressed with the rest of $H_{in}$ and $I$, which adds noise to our measurements. In particular, it could be the case that the target file is in $W'$, but all files in $W''$ are similar to each other. So, if the decrease in size from compressing all files in $W''$ is smaller than the target file, we would recurse into the wrong set. To address this, $\advA$ can measure the compressibility of each injected list of files, and ``penalize'' the measurement according to this. Concretely, on every iteration of the attack, $\advA$ computes $z' = \bigl|\gzip(w_1 \concat ... \concat w_{n/2})\bigr|$ and $z'' = \bigl|\gzip(w_{n/2+1} \concat ... \concat w_{n})\bigr|$ locally, and recurses into $W'$ if \[\obs_2/(\obs_1 + z') < \obs_3/(\obs_1 + z'')\] and into $W''$ otherwise; $\obs_1$, $\obs_2$, and $\obs_3$ respectively denote the encrypted database before both injections, after the first injection, and after the second injection.

After the attack is over, $\advA$ can confirm whether the attack was successful or not by making an additional injection with just the found attachment: if the guess is correct, $\obs$ will have no increase in size (except, potentially, for a negligible number of bytes due to noise from the re-encryptions.)

\paragraph{Attack implementation.}
\definecolor{color1}{HTML}{40B0A6}
\definecolor{color2}{HTML}{E1BE6A}
\begin{figure}[t]
    \small
    \centering
    \begin{tabular}{r|rrrr}
        \toprule
            & \textbf{Enron} & \textbf{$|w|$=10KB} & \textbf{500KB} & \textbf{1MB} \\
        \midrule
            $|W| = 32$ & 92 & 74 & 44 & 47 \\
            $128$ & 81 & 53 & 34 & 38 \\
            $512$ & 52 & 29 & 18 & 8 \\
        \bottomrule
    \end{tabular}
\caption{Experimental success probability of our dictionary attack on attachments exploiting deduplication, for real-world files (first column), and for synthetic test files (each file is of some random size between between 1 and $|w|$, and contains sequences of repeated characters). Each row represents a number of candidate files. These probabilities can be amplified via repetition.}
\vspace{-0.3cm}
\label{fig:attach_dict_experiments}
\end{figure}
We implemented both the naive and the binary-search attacks against KeePassXC, and confirmed that the former successfully recovers the target attachment every time. For the latter, we ran a sequence of experiments to estimate its probability of success. Our experimental setup consisted of two testing environments: (1) using real KeePassXC clients (version 2.7.6), and (2) a local re-implementation of all client-side operations relevant to our attacks, to simulate the real setting.

This dual-setup approach is a common strategy used to evaluate deployed systems~\cite{injectionattackspaper,backendal2023mega, patersonthree}. The real environment allows us to confirm the overall correctness and implementation of our attacks, while the simulated environment allows us to compute an empirical estimate of the probability of success. For the latter, we can run a high volume of trials in a reasonable amount of time, and without overloading the cloud-service provider. In addition, we can ensure that the trials are independent, by using the same starting state across each. We discuss our experimental setup in more detail in Appendix~\ref{subsec:compromised-storage-attacks-details}.

Equipped with this setup, we used three types of datasets to gauge the attack's success rate on different workloads. The first consisted of the Enron corpus~\cite{enron}, which is a public dataset of real-world emails. This dataset helped us evaluate our attack on practical targets. The other two datasets, generated locally by us, consisted purely of synthetic data for benchmarking purposes: a set of random files, all of the same size; and a set of files of varying sizes, and composed of substrings of repeated characters. The reasoning behind these datasets is that random bytes and equal file sizes minimize the effects of gzip noise (since random data is less compressible), while substrings of repeated characters and different file sizes maximize gzip noise.

For each data set, we ran 100 trials of the attack, and recorded the fraction of these in which the target attachment was successfully recovered. Each trial consisted of sampling a fresh set of candidate files of the appropriate type, choosing one file at random, adding it to the victim's vault in a personal entry, and running the attack. For the dataset consisting of files of the same size and all random bytes, our attacks succeed with 100\% probability for all file sizes. The results for the other two datasets are displayed in Figure~\ref{fig:attach_dict_experiments}. As our experiments show, our attack succeeds with high probability. Further, as explained earlier, the adversary can confirm whether the output is correct or not, and re-run the attack if needed until the correct attachment is found, which essentially gets rid of false positives.

\subsection{Leakage from Compression}\label{subsec:leakage-compression}
The second vulnerable mechanism we identified in KDBX~4 is the fact that~$H_{in} \concat I$ is gzip-compressed before it is encrypted. The (oversimplified) way in which gzip works is that repeated substrings are replaced with short pointers to their prior occurrences, if any, which gets rid of redundancy in the payload. The resulting sequence of pointers and (unmatched) characters is then serialized using Huffman coding to yield the compressed bytestring.

Since~$I$ (resp.~$H_{in}$) contains both personal \emph{and} shared entries (resp. attachments), the first step of the gzip algorithm leads to a cross-user interaction that an adversary can exploit: if fields in the injected credentials match some field in~$U$'s personal credentials, the length of the resulting encrypted database will be shorter than if they do not match, as in the former case gzip will replace the injected content with a short pointer to~$U$'s private content.

\paragraph{Attack description: dictionary attack.} Our attack consists of a dictionary attack, i.e., for a list of candidate items~$W \coloneqq \{w_1, ..., w_n\}$, $\advA$ wants to determine which~$w_i$, if any, is in~$\state$. The items of interest consist of login information, except for passwords (e.g., usernames or URLs), and contents of attachments.
The reason why passwords are not recoverable is due to the fact that these are under an additional layer of encryption, which prevents matching passwords from being deduplicated by gzip.

One possible attack approach is to simply inject each candidate string, one at a time, and return the one that leads to the smallest increase in database size. This attack, however, suffers from low accuracy, since it compares compressibility across \emph{different} candidate strings. Since each string contains different characters, the Huffman coding step of gzip may lead to false positives: it could be the case that an incorrect candidate string is composed of characters that appear very frequently in the database, which leads to shorter Huffman codes and the appearance of a more compressible string.

Instead, drawing inspiration from other compression-based attacks such as~\cite{rizzo2012crime,hogan2023dbreach,gluck2013breach}, we use \emph{two} injections per candidate $w$---the first containing $w$ itself and the second a \emph{scrambled} version of $w$---and compare their relative compressibility. Thus, each pair of injections consist of the same characters, which reduces the effects of Huffman coding. (To our knowledge this implementation of the generic ``two-tries'' method is novel.) More concretely, for each $w_i$, $\advA$ first injects $w_i$ through a shared entry $e$ (depending on the type of $W$, e.g., by setting either $e_{url} = w_i$ or $e_{user} = w_i$). Then, $\advA$ records $n_i' = |\obs|$. This is followed by a second injection, where $\advA$ replaces $w_i$ in the relevant field with a random permutation of its characters, and records the updated $n_i'' = |\obs|$. Finally, $\advA$ returns whichever $w_i$ yield the maximum value for $|n_i'' - n'_i|/|w_i|$.

The intuition behind the strategy is that, in the first injection, the correct string gets compressed in its entirety, while incorrect strings only get partially compressed. Then, the second injection serves as a baseline measurement to gauge the ``worst-case'' compression of a string with the same characters as $w_i$, to control for the noise from the Huffman coding. So, it may be the case that some string leads to the smallest total increase in $\obs$ but, in fact, this is also the case when the string is replaced by a scrambled version of it. This confirms that the decrease in size comes from the short Huffman codes and not from redundancy with the rest of the database (otherwise, the second string would ``break'' the effects of compression).

As in the prior attack, $\advA$ can confirm whether the attack was successful or not by, for example, re-injecting the found string, prepended with a pad of 32K bytes (the length of zlib's search window), and confirming that output file increased by the size of the string. 

We note that our dictionary attack can potentially be extended to a \emph{byte-by-byte} recovery attack, where the list of candidate guesses is not known a priori. To do so, $\advA$ can instead employ a CRIME-style attack~\cite{rizzo2012crime}. In such attacks, $\advA$ first has knowledge of a prefix $p$, of length at least three bytes,\footnote{gzip does not compress matching substrings unless they are at least four bytes long.} that precedes the secret, which serves to ``bootstrap'' the attack. To guess the first character, the high-level idea is that $\advA$ tries all possible values $z_i$ for it by injecting $p\concat z_1$ followed by $p\concat z_2$, and so on. Then, all incorrect guesses will get compressed by only $|p|$, but the correct guess will match an \emph{additional} character, leading to a slightly smaller ciphertext. The attacker proceeds in this fashion, one byte at a time, until eventually the entire secret is recovered. If the adversary is extracting information in attachments, the known prefix $p$ can be part of a document's template; for example, if recovering the salary in a contract, $p$ may be ``Salary: ''.  Conversely, if the adversary is recovering URLs or usernames, we can take advantage of the structure of KDBX~4, and set $p$ to be the \emph{XML tags} of the field. For example, if recovering a username, $p = \texttt{<Key>UserName</Key><Value>}$. Further, $\advA$ could recover a field from a \emph{specific} credential, by appending additional information to $p$. For example, to recover the username specifically for \url{target-site.com}, $\advA$ can use $p = \texttt{<Key>URL</Key><Value>target-site.com</Value>}$ $\texttt{...UserName</Key><Value>}$. As before, injections are performed by updating the URL, username, or attachment fields of a shared entry, this time with $p \concat z_i$ as the payloads. Exploring this extension further is outside the scope of our work, as such attacks rely on lower-level details of compression algorithms.

\definecolor{color1}{HTML}{117733}
\definecolor{color2}{HTML}{882255}
\begin{figure}[t]
    \resizebox{\columnwidth}{!}{%
    \centering
    \renewcommand{\arraystretch}{1.1}
    \begin{tabular}{r|rr|rrrr}
        \toprule
            & \textbf{Websites} & \textbf{Usernames} &\textbf{$|w| = $ 5} & \textbf{10} & \textbf{15} & \textbf{20}\\
        \midrule
            $|W| = 4$ & 99 & 80 & \textcolor{color1}{89}, \textcolor{color2}{61} & \textcolor{color1}{100}, \textcolor{color2}{74} & \textcolor{color1}{100}, \textcolor{color2}{86} & \textcolor{color1}{100}, \textcolor{color2}{88} \\
            $10$ & 97 & 66 & \textcolor{color1}{81}, \textcolor{color2}{51} & \textcolor{color1}{98}, \textcolor{color2}{59} & \textcolor{color1}{100}, \textcolor{color2}{57} & \textcolor{color1}{100}, \textcolor{color2}{63} \\
            $25$ & 92 & 54 & \textcolor{color1}{64}, \textcolor{color2}{30} & \textcolor{color1}{97}, \textcolor{color2}{33} & \textcolor{color1}{100}, \textcolor{color2}{48} & \textcolor{color1}{100}, \textcolor{color2}{44} \\
            $50$ & 89 & 46 & \textcolor{color1}{50}, \textcolor{color2}{17} & \textcolor{color1}{93}, \textcolor{color2}{28} & \textcolor{color1}{100}, \textcolor{color2}{29} & \textcolor{color1}{100}, \textcolor{color2}{38} \\
            $100$ & 84 & 24 & \textcolor{color1}{39}, \textcolor{color2}{8} & \textcolor{color1}{92}, \textcolor{color2}{13} & \textcolor{color1}{100}, \textcolor{color2}{16} & \textcolor{color1}{100}, \textcolor{color2}{14} \\
        \bottomrule
    \end{tabular}
    }
    \caption{Experimental success probability of our dictionary attack on usernames and URLs exploiting compression, for real-world files (first two columns), and for synthetic test files (last four columns); the \textcolor{color1}{left} and \textcolor{color2}{right} values denote, respectively, when the strings consist of random bytes and are all of length $w$, and when they have repeated substrings and vary in length between 1 and $|w|$. These probabilities can be amplified via repetition.}
    \vspace{-0.3cm}
\label{fig:url_dict_experiments}
\end{figure}
\paragraph{Attack implementations.}
We experimentally verified our dictionary attack against KeePassXC, using URLs and usernames as examples of items of interest. An additional complication about KeePassXC is that, as explained earlier, every database save rotates $K'$, re-encrypts all passwords, and updates metadata (for example, a timestamp indicating the last modification time of the shared entry). This results in a small number of bytes of noise in the compression side-channel. Given that our target items are only a few bytes long, such as URLs and usernames, this noise is noticeable. So, we use a simple refinement from~\cite{zindros2016practical}: inject each candidate item $t$ times instead of just once, compute the average value of the resulting $|\obs|$, and use this in the measurements. Of course, the larger $t$ is, the higher the probability of success will be. $\advA$ can pick an appropriate value based on the context of the attack (for example, the size of $W$, how long the victim will be using their device, etc).

We tested our attack in the same dual-setup as in Section~\ref{subsec:leakage-attach-dedup} and using the same three types of workloads; the datasets of real-world data consisted of a list of the Majestic ranking of top websites~\cite{majestic-list}, and a corpus of common usernames compiled from various data breaches~\cite{githubcommonpasswords}. The results of our experiments on usernames are displayed in Figure~\ref{fig:url_dict_experiments}, which correspond to the success rate across 100 independent trials of each workload, using $t=10$ for each injection.
\section{Mitigations and Responsible Disclosure}\label{sec:takeaways-and-mitigations}
We discuss mitigations for our attacks in this section, as well as the results of our responsible disclosure with the vendors directly affected by our work. The patches adopted by these may serve as inspiration for other applications that are vulnerable to our attacks.

The most immediate mitigation to our attacks from Section~\ref{sec:leakage-from-security-metrics} would be to remove shared credentials from the computation of vault-health metrics that are logged to the servers, in order to disable the injection side-channel. This, however, would result in a loss of information for users, as they would no longer be able to detect duplicate passwords present in (non-malicious) shared credentials. Other mitigation approaches would depend on the exact purpose of these metrics, which is opaque to us. For example, if the metrics are only used for client-side computation and stored on the server, clients can encrypt these locally before exporting them. Conversely, if the metrics are used to compute server-side statistics, applications could use privacy-preserving aggregate statistics frameworks such as~\cite{corrigan2017prio}.

Our attacks from Section~\ref{sec:leake-url-icons} against a network adversary can be mitigated by fetching icons every time a new credential is added to the vault, even if the URL is a duplicate. Note that clients can potentially still store only one copy of each icon in the client side, and the requests can be repeated just for the sake of hiding fetch patterns. To also hide the URLs from an eavesdropping adversary, applications can use private-information retrieval (PIR) schemes~\cite{chor1998private} to retrieve icons without revealing the URL in question.

A direct mitigation to our attacks from Section~\ref{sec:compromised-storage-attacks} is for applications to, of course, disable storage-saving mechanisms. However, this could result in a prohibitive blowup in their storage footprint. By definition, disabling deduplication doubles the cost of storing repeated attachments. The cost of disabling compression depends on the file format and the underlying user data; for highly structured file formats like XML, the increase is particularly noticeable. Local tests on example KeePassXC databases resulted in 5-6x increases in size.

Another mitigation approach is to confine the storage-saving mechanisms to subsets of the application data that are within the same trust context, which would disable the injection channel. For deduplication, this translates to deduplicating files separately for every shared folder. More work is required to devise an analogous solution for compression. A different mitigation approach for compression would be to isolate sensitive fields within the database by encrypting them under a second layer of encryption, which prevents similar data from matching with these fields. This is analogous to how KeePassXC protects passwords, as described in Section~\ref{sec:compromised-storage-attacks}, and indeed they could consider encrypting other fields in a similar fashion. Yet another approach could be to add padding or noise to the database before encrypting it, in order to obfuscate the real size of the database. Adversaries can use statistical techniques to adapt to such mitigations, however, as has been the case in other contexts such as network traffic fingerprinting (see, e.g.,~\cite{dyer2012peek}).

\paragraph{Responsible disclosure.}
We reported our findings to the ten vendors affected by our work, many of which proceeded to deploy mitigations.

LastPass adopted our suggested mitigation of separating vault-health metrics between personal and shared credentials, which disables the injection channel. They released an initial implementation of this fix in version 4.129.0, removing shared folders from the vault-health logs, as these lead to the most severe variant of our attack. Removing individually shared credentials from the logs is more technically challenging---and individual credentials lead to a less practical version of our attack---and thus has been deferred to later in their roadmap; their projection is to release this fix by the end of the year, which would complete a full mitigation to our attack.

Zoho Vault plans to adopt a similar fix, by implementing an option to separately compute vault-health metrics on personal passwords as of version 4.0. Dashlane opted for a partial mitigation instead, namely, increasing their rate limits on the sharing endpoints ``as much as possible''. Given the fact that their vault-health metrics are only logged once per day, their tight sharing limits significantly affect the practicality and runtime of the attack. In addition, the resource limits on their web application and extensions prevent an adversary from sharing an unlimited number of credentials with a victim, which increases the runtime of the attack even more. As part of the disclosure, they informed us that incorporating shared passwords is a core feature of their vault-health metrics, and thus removing shared passwords would represent a notable disruption to this feature. 

Then, to address our URL icon fetching attack, Dashlane implemented a new feature as of version 6.2415 that allows users to turn off fetching credential icons, which disables the side-channel for both an eavesdropping and network adversary, and thus provides a full mitigation to our attack; details of this new feature can be found at \cite{dashlane-url-icons-new-feature}. To address our attack even when URL icons are turned on, they additionally migrated their icon fetching tool to a new endpoint (\url{api.dashlane.com}), which is used by multiple parts of their application logic. As such, this would make it significantly more challenging for a network adversary to use traffic analysis techniques to identify whether an icon fetch request is included in the traffic sent to this top-level endpoint, due to the high amount of noise from the other requests sent to this endpoint.

NordPass also adopted our suggested mitigation of separating vault-health metrics between personal and shared credentials, which disables the injection channel and is thus a full mitigation to our attack. This fix is planned to be deployed by the end of August 2024. Then, to address our URL icon fetching attack, NordPass added a feature to disable URL icons by default, which provides a mechanism to disable the injection side-channel. This fix is planned to be deployed before the end of 2024. In addition, they are currently exploring more robust mitigations for this attack, to protect users even when URL icons are turned on.

To address our attack on attachment deduplication, KeePassXC adopted our suggested mitigation of deduplicating files separately for every shared folder, which disables the injection side channel. Then, to address our compression-based attacks, they modified their file format by, every time the database is saved, picking a random length between 64 and 512 bytes, generating a random array of this length, and including this in a ``custom data'' field of their file format. We note that this is only a partial mitigation, as an adversary can potentially use statistical techniques to bypass the noise; this, however, would require a significantly higher number of injections. Both fixes were promptly implemented by the KeePassXC team, and have since rolled out as part of version 2.7.

Enpass already provides support for turning off URL icons (which is off by default)~\cite{enpass-rich-icons-off}, and thus users who disable URL icons are not vulnerable to our attack. As a first step towards more mitigations, Enpass added an option for organization admins to control this setting via an organization level policy. They have decided not to deploy mitigations at this time to address the attack even when URL icons are turned on.

1Password already provides support for turning off URL icons~\cite{1password-rich-icons-off}, and thus users who disable URL icons are not vulnerable to our attack. However, 1Password decided not to deploy additional mitigations to address the attack at this time, even when URL icons are turned on. Similarly, Proton Pass already provides support for turning off URL icons, and thus users who disable URL icons are not vulnerable to our attack. We shared suggestions for how to address this attack even when URL icons are turned on, but do not have details on their plans to deploy these.

Lastly, Keeper considers our attack on their system a very low severity issue and opted not to deploy mitigations, and instead have added changing this feature as a consideration for an upcoming platform update. As part of the communication, they shared that removing transferred credentials from the count of duplicate passwords displayed in the Admin Console would represent a notable disruption to a feature of the business product.
\section{Conclusion}\label{sec:conclusion}
We introduce a new threat model for password managers in this work, which we exemplify via four general attacks, using ten applications as case studies. Our attacks suggest the need to rethink certain aspects of password manager design, and of E2EE applications more broadly. We highlight some takeaways in this section.

Our attacks from Section~\ref{sec:leakage-from-security-metrics} are symptoms of the more general pattern of exporting application state that is a function of both personal \emph{and} externally-received data, which can potentially open the door for injection attacks. Thus, a guiding principle for E2EE application designers is to \emph{separate data according to the trust assumptions of their system}. Our attack from Section~\ref{sec:leake-url-icons} is an example of the broader pattern of fetching content from external sources in a \emph{state-dependent way}, which an adversary can potentially exploit by injecting payloads and seeing how this affects subsequent fetches. Thus, a guiding principle for application designers is for client-server communication to be \emph{as resource-specific as possible, and to not depend on the results of prior operations}. Lastly, our attacks from Section~\ref{sec:compromised-storage-attacks} serve as an example of the tensions between security and performance. Since storage-saving mechanisms get rid of redundancy in application files, techniques of this form naturally pave the way for potential injection attacks. As such, more work is required to understand how to balance storage costs and security, particularly in the context of compression and file deduplication, and to devise frameworks that help explore these trade-offs in a principled way.

We disclosed our attacks to the vendors affected by our work, many of which deployed fixes to address these. These mitigations, however, are bespoke solutions for the attack vectors presented in this work; a central direction for future research is to design general-purpose detection and mitigation techniques against injection attacks, and to deepen our understanding of this threat model.

\section*{Acknowledgements}

This work was supported in part by NSF 
grants CNS-1704296 and CNS-2120651.

\bibliographystyle{plain}
\bibliography{reference}

\appendix
\section{Testing Methodology and Analysis}\label{sec:methodology}
We explain the testing methodology for our attacks in this section, where we include a summary of the steps shared with vendors that allowed them to reproduce the attacks.

\paragraph{General setup.} For each analyzed application, we first created two separate accounts for the service. The devices used were laptops running macOS Monterrey. If relevant for the application, we additionally created an organization under which both accounts were registered. Then, we established a shared folder (or individual credentials, if the application does not support folders) between both accounts.

\subsection{Attacks from Sections~\ref{sec:leakage-from-security-metrics} and~\ref{sec:leake-url-icons}}
To simulate the view of both a compromised server and network adversary, we used Charles Proxy~\cite{charlesproxy} to establish a proxy server on the victim's device. This allowed us to intercept both the encrypted HTTPS traffic, and enable SSL proxying to intercept the unencrypted traffic that would reach the application servers.

Our testing data consisted of an open-source list of common passwords~\cite{githubcommonpasswords} and the Majestic ranking of top websites~\cite{majestic-list}, for the attacks in Section~\ref{sec:leakage-from-security-metrics} and~\ref{sec:leake-url-icons}, respectively. We explain the main steps for our attack from Section~\ref{sec:leakage-from-security-metrics} below; the steps for the attack in Section~\ref{sec:leake-url-icons} are analogous, except that we inject URLs instead of passwords.

To implement the attack, we first sampled uniformly at random a password from this list, and added it to the victim's vault as the target password. We then sampled at random various subsets of passwords of increasing sizes (1 through 512, in increasing powers of 2), which represented the list of candidate passwords. Lastly, we ran the attack from the adversary's account, following the steps described in Section~\ref{sec:leakage-from-security-metrics}. At a high-level, this consisted of creating a batch of shared passwords in the adversary's account, waiting for the next request from the victim's account reporting the updated number of duplicates (which was intercepted by our proxy server), and repeating these steps, until completion of the attack.

To determine the network delay for each log of duplicates, as reported in Section~\ref{sec:leakage-from-security-metrics}, we compared (1) the timestamp from the initial request for the upload of the shared credentials, against (2) the timestamp of the request that logged the number of duplicate passwords. We computed the average delay between each injection. Then, the total time to run each attack is simply $t \cdot n$, where $t$ is the number of iterations the attack requires, and $n$ is the experimental delay between logs of duplicates.

Both attacks succeed with 100\% probability by design: computing the number of duplicates in the vault and fetching URL icons are both deterministic, noise-free operations. Thus, there was no need for an experimental estimate of the attack's correctness. Indeed, our attack implementations succeeded every time.

\subsection{Attacks from Section~\ref{sec:compromised-storage-attacks}}\label{subsec:compromised-storage-attacks-details}
To simulate the view of a compromised encrypted vault, it sufficed to place the encrypted database file in a local folder in the victim's device, and measure its size directly.

We verified the general correctness of the attacks by implementing them using real KeePassXC clients, and then computed an empirical estimate of their probability of success on a local simulation (described below). We used various data sources as testing data for the attacks, as described in Sections~\ref{subsec:leakage-attach-dedup} and~\ref{subsec:leakage-compression}. For each, we first added the target item to the victim's device. Then, we ran the attack from the adversary's device as explained in the relevant sections, by iteratively adding credentials to the shared folder with injected items and measuring the updated size of the victim's encrypted vault. In this way, we were able to confirm the presence of the deduplication and compression mechanisms.

\paragraph{Statistical analysis.} We implemented a local simulation of the core operations of the KeePassXC client that are relevant to our attacks, which allowed us to compute the statistical error rates of our attacks. Concretely, our simulation received as input an initial KDBX 4 database, and provided an interface for adding shared credentials to the database, which would update the vault in-place, in a way that is consistent with updates from real clients. This consisted of two steps: (1) Modifying the plaintext database, by adding the shared credential as a new XML element to the payload, adding new attachments to the inner header (if any), and updating metadata; and (2) Re-encrypting the new plaintext database, by rotating the secret key in the inner header, and compressing-then-encrypting the inner header and XML payload with this key.

Our implementation of this simulated environment was guided by KeePassXC's open-source codebase, in addition to manual comparison of decrypted KDBX 4 databases before and after a shared credential was added, which revealed all modifications that resulted. In particular, this is derived from files~\url{src/format/Kdbx4Writer.cpp} and~\url{src/format/KdbxXmlWriter.cpp} in KeePassXC's codebase~\cite{keepassxc-code}. We tested the consistency of this simulation by adding shared credentials via both our local implementation and real KeePassXC clients, and verifying that the resulting database states are equivalent in both cases.

Equipped with this simulation, we ran the attacks (following the steps described in Section~\ref{sec:compromised-storage-attacks}) in a single device, to be able to compute a high number of trials without overloading the application servers. We used the interface provided by our simulation to add credentials to the vault (which mocks the injected credentials), followed by local measurements of the resulting file size (which mocks the adversary's view of the compromised encrypted vault). This allowed us to empirically estimate the noise present in our attacks, by repeating each experiment many times on the same starting state, and tallying the number of successful attempts (see Section~\ref{sec:compromised-storage-attacks} for more details, and the reported error rates).

\section{Details of Case Studies from Section~\ref{sec:leakage-from-security-metrics}}\label{sec:leakage-metrics-attacks-details}
We explain additional details of the applications for which we deployed proof-of-concept implementations of our attack from Section~\ref{sec:leakage-from-security-metrics} in this section.

\paragraph{Case study \#1: LastPass.} LastPass's Chrome extension and web vault log vault security metrics via POST requests to the endpoint \url{lastpass.com/lmiapi/users/me/security/score}.\footnote{This is only the case for users with premium accounts (any one of Families, Teams, Premium, and Enterprise plans, and the trial versions of these).} The payload of this request, among other things, maps the string ``$\texttt{numallduppasswords}$'' to the number of duplicate passwords in the vault, i.e., $\dup(\state)$. Importantly, this metric only quantifies how many passwords themselves are reused, and not across how many accounts. That is, if there are already two copies of a password in a vault, adding a third copy does not increase this metric. As such, our attack on LastPass requires the additional assumption that $U$ only has one copy of the target password in their vault at the start of the attack.\footnote{Similarly, the basic attack requires a trivial refinement: if $P_{found} = P'$ in step (5), $\advA$ needs to first delete or overwrite the entries they just injected, as otherwise there would already be a copy of the target password in the vault (from their prior injection).}

The POST requests to \url{lastpass.com/.../score} are triggered after every modification to any entries in $\state$. In particular, a request is sent as soon as $U$'s client automatically imports the updates to the shared folder $F$. Synchronization of changes to shared folders is very frequent. Indeed, LastPass claims that changes are ``synchronized automatically and propagate to everyone with whom the folder has been shared''~\cite{lastpasssharedsync}. In practice, there is a short time delay, which represents the time the adversary has to wait between each of the $\lceil \log_2 n \rceil$ injections. We measured this to be between 3 to 5 minutes on average.

Besides accepting the shared folder before the attack starts, our attack requires no interaction from $U$, and only requires $U$ to be logged into their account (for example, in their Chrome extension). Importantly, the victim need not have
their vault open while the attack runs: if a user is logged in,
a background script checks for updates and syncs them to
the vault, which is enough to trigger the requests. As such,
the attack can operate silently in the background while U is
engaged in other tasks.

\paragraph{Case study \#2: Dashlane.} Dashlane's web vault and Chrome extension log their vault security metrics via POST requests to \url{styx.data.dashlane.com/v1/event/user}. The payload of this request contains the string ``$\texttt{passwords\_reused\_count}$'', which corresponds to $2 \cdot \dup(\state)$. Unlike LastPass, this metric quantifies the number of \emph{accounts} with reused passwords, i.e., we do \emph{not} need to assume that $U$'s vault only contains one copy of the target password. More generally, the attack requires no assumptions about the (lack of) activity by $U$,\footnote{Unlike LastPass, when the metric increases $\advA$ can distinguish false positives from legitimate matches by unsending the candidate, and seeing if the metric decreases again.} and thus falls under the noisy device setting.

Updates to shared items are immediately synchronized across vaults by background scrips. However, unlike LastPass, the POST requests to \url{styx.data.dashlane.com/.../user} are not triggered automatically after a sync. The exact periodicity of these requests is opaque to us, but we noticed they are triggered once per day (or the next time $U$ logs into their device, if more than 24 hours have passed).

Dashlane does not support shared folders, and only allows users to share individual entries. As such, $\advA$ must use the sequential injection strategy. We note, however, that Dashlane will soon support shared folders: on a July 2023 blog post~\cite{dashlanesharedfolderssoon}, they state that this feature is ``coming soon''. This would thus enable our more efficient binary-search attack variant.

These two limitations make the attack on Dashlane more time-consuming than against LastPass: the adversary needs to perform $\frac{n}{t} + \lceil \log_2 t \rceil$ injections, interleaved between daily requests from $U$ to \url{styx.data.dashlane.com/.../user}. However, since our attack is robust to external noise, $\advA$ can just silently run an extended attack, irrespective of the activity of the victim, until the password is found. As before, the attack only requires the victim to be logged into their account on their browser extension or the website: the synchronization of injections and daily logs of vault metrics are all handled by background scrips.

\paragraph{Case study \#3: Zoho Vault.} Zoho Vault's web vault logs their vault security metrics via POST requests to the endpoint \url{vault.zoho.com/api/rest/json/v1/dashboard/sendAssessmentDetails}. The payload of this request contains an array $O$ of JSON objects, with the format described in Section~\ref{subsec:case-studies-leakage-from-security-metrics}. By counting the number of objects such that ``$\texttt{reused}$'' maps to $\true$, the adversary can identify $\dup(\state)$, and thus run the attack using the binary-search injection strategy. However, Zoho Vault is revealing much more granular information: $\advA$ learns whether the password of each \emph{specific} entry is a duplicate or not, instead of just the total number of duplicate passwords. So, the adversary can simply inject all candidate passwords in a \emph{single} batch, and look for the $\texttt{reused}$ field of the objects in $O$ that correspond to the injected credentials.

The challenge with this approach is identifying which specific objects correspond to which specific injected credentials. To do so, the adversary can leverage an additional fact: upon creation, each credential is assigned a unique string identifying it, which is also stored in each credential's object in $O$. This ID is preserved after a credential is shared. So, the adversary can (1) locally create a credential for each candidate password, (2) note the assigned ID for each, and (3) look for these IDs in $O$ after sharing the credentials. This allows the adversary to map every injected credential to its corresponding object in $O$, as desired.

Unlike the prior attack on LastPass, the correctness of this attack does not depend on $U$ having only one copy of the target password in their vault. Further, our attack is robust to any kind of external noise, and thus falls under the noisy device threat model.

The downside of this attack is that requests to \url{vault.zoho.com/.../sendAssessmentDetails} are not very frequent; in particular, unlike LastPass, they are not triggered automatically after every injection. Instead, they are triggered when $U$ logs into their Zoho Vault account, if enough time has passed after the prior security report was sent; empirically, we noted this gap to be roughly once per day. In addition, $U$ can manually trigger these requests if they refresh their password security metrics in the GUI of the web application. In any case, the adversary can just run the first three steps of the attack, and wait for the next request to be triggered (at most 24 hours). We stress that the adversary only needs to see \emph{one} request to this endpoint, so the adversary only needs to wait once.

To extend this attack to a network adversary, we can leverage the ideas discussed in Section~\ref{subsec:case-studies-leakage-from-security-metrics}, as follows:

\begin{newenum}
    \item \emph{Setup:} Create $P/2$ empty, shared credentials.
    \item \emph{Baseline measurement:} Wait for an HTTPS request from $U$ to $S$ that logs the vault-health metrics, and record the size of the (encrypted) payload $T'$. Let $n_1 \coloneqq |T'| = |O|$. Note that $O$ will contain an object for each of the credentials created in the preceding step, and such that their $\texttt{reused}$ flags map to $\false$.
    \item \emph{Inject:} Split $P$ into two halves $P', P''$ of equal size. Assign each $p \in P'$ to the password field of each of the empty credentials from step (1).
    \item \emph{Measure:} Wait for another HTTPS request from $U$ to $S$ that logs the vault-health metrics, and record the size of its encrypted payload $T''$ once again, which corresponds to $n_2 \coloneqq |T''|$.
    \item \emph{Recurse:} If any of the injected passwords from step (3) matched some victim password, its corresponding object in $O$ would \emph{flip} its $\texttt{reused}$ flag from $\false$ to $\true$ (while the rest of the metadata remains unchanged), which would imply that $n_2 < n_1$ by at least one byte. If so, recurse into $P'$, and into $P''$ otherwise, until there is only one candidate password left.
\end{newenum}

Since the attack requires very precise measurements, we must assume that the device is free of external noise at every recursive round, in order for fluctuations in $|O|$ to correspond only to changes in the $\texttt{reused}$ flags of injected passwords. Note, however, that the device need not be quiet for the \emph{entire} attack: if $\advA$ notices that external changes also took place, they can simply repeat this round of the attack. As such, the attack as a whole falls under the noisy device setting.
\section{Details of Case Studies from Section~\ref{sec:leake-url-icons}}\label{sec:leakage-url-icons-details}
We explain additional details of the application for which we deployed proof-of-concept implementations of our attack from Section~\ref{sec:leake-url-icons} in this section.

Adding a new credential in Dashlane results
in a POST request to \url{ws1.dashlane.com/2/iconcrawler/getIcons}, which retrieves the endpoint where the corresponding URL icon is stored. This is followed by a request to an endpoint of the form \url{d2erpoudwvue5y.cloudfront.net...} (determined by the first request) to actually fetch the icon itself. We note that the second request requires the victim to have their vault open (instead of just being logged in), whereas the first one does not.

The payload of the request to \url{ws1.dashlane.com/2/iconcrawler/getIcons} contains the URL in plaintext,
which trivially leaks to an eavesdropping adversary all the
URLs in a victim's vault. As mentioned in Section~\ref{sec:leake-url-icons}, however,
the fact that icons are only fetched once (even if a URL is added under multiple credentials) can be leveraged by a network adversary to infer whether a particular candidate URL $w$ is in a victim's vault or not: they can inject a credential
with w as the URL, and inspect whether an icon fetch request
is triggered from the victim's client or not. In this case, the
plaintext payload of the request is not required, hence why a
network adversary is sufficient.

If the victim has their account open, the resulting request (or
lack thereof) to \url{d2erpoudwvue5y.cloudfront.net} denotes whether $w$ is a reused URL or not. If the victim has their vault closed, the first request to \url{ws1.dashlane.com/2/iconcrawler/getIcons}, which is triggered automatically, can be used instead (or, alternatively, the adversary can wait until the victim opens their vault). A challenge with this approach, however, is that a network adversary only sees the top level end-point, i.e., \url{ws1.dashlane.com}, and
full request path is encrypted. Since syncing a new credential
triggers more request to this endpoint, the adversary cannot immediately tell whether a request corresponding to the specific path \url{2/iconcrawler/getIcons} was triggered or not as part of all resulting requests to \url{ws1.dashlane.com}.

To work around this, however, the adversary can simply use
standard traffic analysis and website fingerprinting techniques.
For example, they can first inject some unintelligible, random
string $\hat{w}$ as the URL, and measure the URL of the request and
response payloads transmitted to/from \url{ws1.dashlane.com}.
Since $\hat{w}$ is guaranteed to not be in the victim's vault, there will
be a request to \url{2/iconcrawler/getIcons} as part of all the requests directed to \url{ws1.dashlane.com} as a result of adding this new credential. As such, this serves as a baseline to compare future requests sizes: if injecting $w$ results in payloads of a similar size, this implies that an icon fetch request was also
triggered; conversely, if the payloads are noticeably smaller,
this implies that a request to \url{2/iconcrawler/getIcons} was
not triggered.

We note that there are a few edge cases that lead to false
positives in this attack. First, if the victim had the target URL
in their vault, but deleted it not too long ago, the icon is still
cached by the client, and as such it will not fetch an injected
URL, even if it is no longer present in the victim's vault. In
any case, this still leaks to an adversary that the victim had
an account with this website not too long ago. Second, if the victim adds new credentials the moment when the adversary injects a new candidate URL, the adversary will not be able to tell whether the requests to fetch icons correspond
to their injected credentials or to the victim's, since all they can observe are packets being sent to the aforementioned endpoints. To circumvent this, the adversary can time their injections appropriately.

\end{document}